# Asymmetric Collision of Concepts: Why Eigenstates Alone are Not Enough for Neutrino Flavor Oscillations*

by John Michael Williams

jwill@AstraGate.net
*Markanix Co.*
P. O. Box 2697
Redwood City, CA 94064

\* Preprinted as *arXiv* `physics/0007078`

# 1. Abstract

The symmetry of the problem of the apparent deficit in upward-going atmospheric muon neutrinos reveals two possible, nonexclusive kinds of solution: Nonlinearity in distance or nonlinearity in angle of observation.

<u>Nonlinearity in distance</u> leads to the most popular theory for the atmospheric problem, neutrino flavor oscillations. If the observed deficit is caused by oscillations and not, say, flavor-changing or other weak-force scattering, neutrinos must be massive. But, if flavor oscillations occur in vacuum, all oscillating neutrinos must have exactly equal mass. Theories of oscillation in matter such as the Mikheyev-Smirnov-Wolfenstein (MSW) effect do not work in vacuum. This is the conceptual conflict of kinematics versus vacuum oscillations.

Flavor-changing oscillations like those of the Cabibbo-Kobayashi-Maskawa (CKM) quark theory become possible in vacuum if freely propagating neutrinos may be associated with local substructure.

<u>Nonlinearity in angle of observation</u> leads to a simple prediction of an excess of horizontal muon flavor. This and other angle-based effects should be observable at Super-Kamiokande or other instruments which can measure atmospheric flux by flavor.

> Note: This Abstract was the text on Poster 0. The other posters shown, with some minor corrections, are included in this paper as numbered graphics in blue frames.

# 2. Atmospheric Muon-Neutrino Deficit

Briefly, the deficit is that only a fraction of the muon neutrinos ($n_m$) assumed created in the atmosphere on the opposite side of the Earth are detected. Using a Monte Carlo simulation to factor out instrumental systematics and local geographic differences at Super-Kamiokande, the ratio of expected $n_m$ flux from the far side of the Earth *vs* that simulated is about 2:3. So, a large fraction of the $n_m$'s assumed passing through the Earth can not be detected on the far side. Electron neutrinos ($n_e$) do not seem to be affected by the Earth to the same extent as $n_m$'s.

The $n_m$'s and $n_e$'s are believed almost all to be created because of pion-decay events. Pions are created in the atmosphere by scattered cosmic particles, mostly protons or heavier nuclei. Ignoring here the difference between particles and antiparticles, each pion decays almost immediately into a muon and a $n_m$; the muon then decays (usually in the atmosphere) into an electron, producing one $n_m$ and one $n_e$. So, assuming that a $n_m$ produced by pion decay is the same particle as one produced by muon decay, the initial atmospheric neutrino flux would be expected to consist of about two $n_m$'s for every $n_e$. This expectation is theoretically well supported and has been confirmed within a few percent [2, 3]. But, $n_m$'s from the atmosphere overhead at Super-K are more numerous than those arriving upward through the Earth.

A neutrino oscillation theory attempts to account for the apparently missing upward $n_m$'s by postulating that they have changed reversibly ("oscillated") into neutrinos of some other type. An apparent deficit in $n_e$'s created by nuclear reactions in the Sun also is observed, and it requires yet other theoretical oscillation parameters to explain it; but, this solar disappearance problem will not be discussed here. Several reactor-based and astronomical studies are in progress, and the empirical bounds on the strange behavior of neutrinos should become much clearer in a year or two.

But, the theory is the problem at hand. The present paper is dedicated to an analysis of apparent errors in the usual statement of neutrino oscillation theory, and to an attempt to make it work.

## *2.1 Linearities and Integration-Ratio Parameters*

We look closely at the theoretical context in which the atmospheric deficit in upward-going muon-neutrinos is identified, trying to understand all the relevant parameters. We use an approximate representation of the Super-Kamiokande neutrino telescope (Super-K) as a basis of calculation.

### 2.1.1 Density Function for the Earth's Atmosphere

A generic approach to the Earth's atmosphere just requires integration over the volume of a thin shell. For example, to compute the mass $M$ of the atmosphere, define coordinates as in Fig. 1:

Let the Earth have an average radius of 6400 km, and the atmosphere a density space constant of 4 km. An element of atmosphere located at point $\vec{p} = p(r, \mathbf{q}, \mathbf{f})$ will be associated with a density $\langle \vec{d} \rangle$ of $d(r) = K e^{\frac{R-r}{k}}$, in which $K$ is the average density at sea level (about $1.3 \text{ kg}/\text{m}^3$). For the mass $M$ of the atmosphere, we then have,

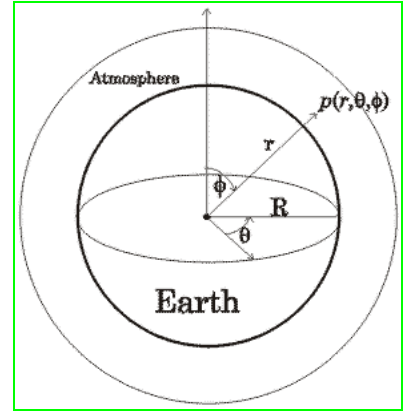

**Figure 1. Coordinates of a point $p$ in the Earth's atmosphere.**

$$M = \iiint \vec{d} = \int_0^{2\pi} d\mathbf{q} \int_0^{2\pi} d\mathbf{f} \int_R^{\infty} dr (\sin \mathbf{f}) r^2 K e^{\frac{R-r}{k}} \text{ ; or,} \tag{2.1}$$

$$M = K \int_0^{2\pi} d\mathbf{q} \left[ 4 \int_0^{\frac{\pi}{2}} d\mathbf{f} (\sin \mathbf{f}) \right] \int_R^{\infty} dr\, r^2 e^{\frac{R-r}{k}} . \tag{2.2}$$

After some arithmetic, $M = 5.3 \cdot 10^{18}$ kg, agreeing reasonably well with the known value [1] of about $4.5 \cdot 10^{18}$ kg. Thus, we validate our density function to be,

$$d(h) = Ke^{\frac{-h}{k}} = (1.3)e^{\frac{-h}{4 \cdot 10^3}} \text{ kg}/\text{m}^3 \text{, for } h \text{ the altitude in meters.} \qquad (2.3)$$

### 2.1.2 Atmospheric Neutrino Observation Parameters

We assume as an approximation that the probability of showering of a primary cosmic particle depends only on density of the atmosphere; the probability of a subsequent decay then will depend on the inverse density [2]. For any such event, there will be a flux of (anti)neutrinos with probability of emission $P_N$ of a neutrino of type $N$, $N \in \{e, \mu, \tau\}$. We consider this flux as normalized against the flux of just one species, say $N = electron$.

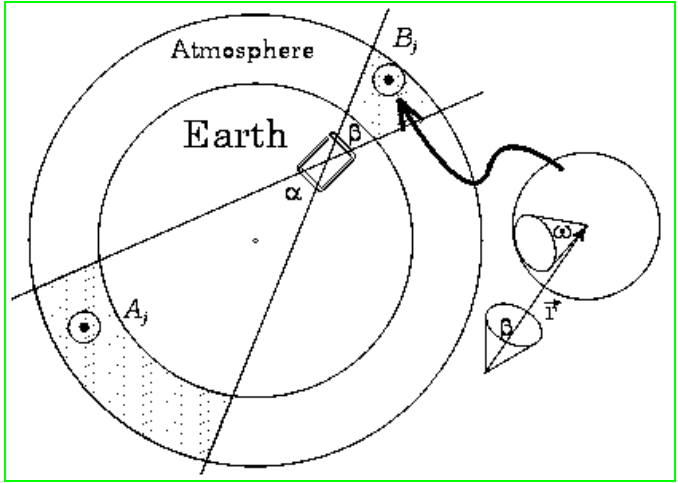

Consider the atmosphere of the Earth a thin shell on a sphere. Consider the observed deficit the result of comparison of counts against some benchmark particle (maybe $n_e$) in two opposite solid angles $a$ and $b$: The two angles may be assumed to define two cones of observation each with apex at Super-K, the base of one directed skyward for downward events ($b$, Fig. 2), the other toward the center of the Earth for upward events ($a$, Fig. 2). Here, $a$ means "up" and $b$ means "down", referring to the observed direction of propagation of the decay products.

**Figure 2.  Analysis of neutrino flux from atmospheric decay events.  For two solid angles (cones) of observation, $a$ and $b$, the shaded regions show the atmosphere in the cones.  Super-K is represented by the rectangle.  Typical events $A_j$ or $B_j$ are assumed to create neutrinos with intensity $I_j(w_j)$, the distribution over solid angle $w_j$ depending on the decay kinematics.  The local coordinate detail for the element $A_j$ is the same as for $B_j$, except for the opposite direction of $\vec{r}$.**

Obviously, letting $a = -b \Rightarrow |a| = |b|$, we immediately expect the upward $\Phi_a$ and downward $\Phi_b$ fluxes to be equal for a thin-shell atmosphere, because the linear extents of the thin-shell areas swept by equal angles at Super-K always will be proportional, and the linear distances to the thin shells always will be inversely proportional.

Nevertheless, we seek more insight into the details. In Fig. 2, the flux $\Phi_b$ in cone $b$ will be assumed given by downward decay events $B$ inversely proportional to density $d(h)$ of the atmosphere. After some figuring, using orthogonal components $a_1$ and $a_2$ to define a rectangular-based pyramid approximation to the differential circular cone angle $a$ in Fig. 2, and likewise for $b$, we get an expression for up-down bias, assuming isotropic scattering in the atmosphere,

$$bias_{iso} \equiv \frac{\Phi(a)}{\Phi(b)} = \frac{\int_{a_1} da_1 \int_{a_2} da_2 \int_{r_{up}^{min}}^{r_{up}^{max}} dr (\sin a_1)\left(d\left((r+D^{min}-2R)\cos a_2\right)\right)^{-1}}{\int_{b_2} db_1 \int_{b_2} db_2 \int_{r_{down}^{min}}^{r_{down}^{max}} dr (\sin b_1)\left(d\left((r+D^{min})\cos b_2\right)\right)^{-1}} . \quad (2.4)$$

We shall not pursue this computation further here, but we see that (2.4) represents purely a geometric ratio, again having nothing to do with particle kinematics or detector energy sensitivity. Taking equal observation cones $a = b = h$ on a diameter of the Earth, and using the exponential atmosphere as in (2.3), we may evaluate (2.4) easily. The cone angle integrals on $a_1$ and $b_1$ drop out when the angles are equal, and the $\cos(h)$ terms may be neglected in the exponential-atmosphere density terms, allowing the second angle integrals to drop out, too. The result,

$$bias_{iso} = \frac{\int_{r_{up}^{min}}^{r_{up}^{max}} dr\left[d(r+D^{min}-2R)\right]^{-1}}{\int_{r_{down}^{min}}^{r_{down}^{max}} dr\left[d(r+D^{min})\right]^{-1}} , \quad (2.5)$$

again shows the same linearity as was evident in Fig. 2.

Notice that it is only the limits of integration that matter in this analysis. So, were further event-locus elaboration attempted, say, by including additional internal, concentric thin shells and expressions for upward decay of ultrahigh-energy secondary particles, the result based on the normalized $P_N$ would not change.

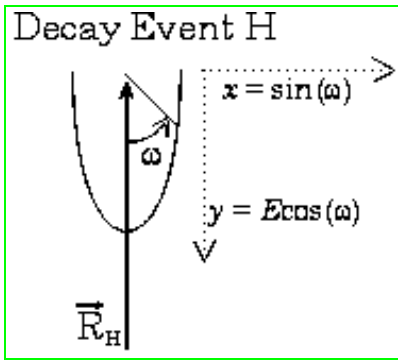

**Figure 3. Distortion of a sphere to represent forward scattering with anisotropy $E$.**

The only parameter missing from (2.5) is that of the scattering profile of the cosmic particles. To describe approximately the forward-scattering intensity, we may write the equation of a circle (sphere) centered on the event $H$ and then distort it by stretching it along a radius $R_H$ of the Earth, as in Fig. 3, with the eccentricity parameter $E$. Symmetry about $R_H$ makes the signs of the angles unimportant.

We obtain our result then by expressing $w$ in terms of $a_2$ and modifying Eq. (2.5); the bias numerator gains a scattering-intensity term so that,

J. M. Williams    2001-08-14 v. 1.6    6

$$\Phi_{up} = \int_{\mathbf{a}_2} d\mathbf{a}_2 \int_{r_{up}^{\min}}^{r_{up}^{\max}} dr \left[ d(r + D^{\min} - 2R) \right]^{-1} I(\mathbf{a}_2); \qquad (2.6)$$

and likewise for the bias denominator. Using Fig. 3 to specify the intensity function, we may write,

$$bias_{\text{aniso}} = \frac{\Phi_{up}}{\Phi_{down}} = \frac{\int_{\mathbf{a}_2} d\mathbf{a}_2 \int_{r_{up}^{\min}}^{r_{up}^{\max}} dr \left[ d(r + D^{\min} - 2R) \right]^{-1} \left( \sin^2 \mathbf{a}_2 + E^2 \cos^2 \mathbf{a}_2 \right)^{\frac{1}{2}}}{\int_{\mathbf{b}_2} d\mathbf{b}_2 \int_{r_{down}^{\min}}^{r_{down}^{\max}} dr \left[ d(r + D^{\min}) \right]^{-1} \left( \sin^2 \mathbf{b}_2 + E^2 \cos^2 \mathbf{b}_2 \right)^{\frac{1}{2}}}. \qquad (2.7)$$

However, once again, the integration over equal cone angles would make all angle terms drop out, and the scattering profile would make no difference. So, without prejudice, we also have confirmed that the atmospheric neutrino flux should obey optical conservation of étendue (optical generalized Lagrange invariant) [4, I.1.6], which corresponds to equality of the angle of observation integrals.

The only relevant parameters therefore are <u>distance of the detector</u> from the detected atmospheric scattering event and <u>angle of observation</u>. There seems not to be any way that the ratio over equal observation angles can depart from the ratio determined by the cosmic particle decay event probabilities $P_N$, for example, $P_{n_m}/P_{n_e}$, unless there is allowed some factor nonlinear in these parameters.

Therefore, any theory explaining the observed deficit must postulate or imply a nonlinearity in observation distance or in observation angle or both.

### 2.1.3 Atmospheric Neutrino Observations

The spherical Earth and its isotropic atmosphere above of course are an idealization. Actual data would dictate the degree of asymmetry and therefore empirical nonlinearity in the real Earth and its atmosphere. In particular, the cosmic charged-particle flux is very strongly influenced geographically by the Earth's symmetry-breaking magnetic field [17]. At least one Monte Carlo model of charged-particle interaction with the Earth's magnetic field [14] has shown that the flux of muons at Super-K would be weakened merely by the magnetic field to about the observed value of neutrino disappearance. Of course, one reason the Super-K atmospheric data are given in terms of a double ratio, $\left( \Phi_{up}/\Phi_{down} \right)_{\text{observed}} / \left( \Phi_{up}/\Phi_{down} \right)_{\text{monte carlo}}$, is to factor out this geographic effect.

## *2.2 Two Nonlinearities*

### 2.2.1 Distance

#### 2.2.1.1 Oscillations

Several derivations of the flavor-distance formula have been given in [5], Sections 7.2 - 7.4. We reproduce one of them here to show the major factors in the derivation for vacuum oscillations. The result for MSW oscillations in matter is similar, but the MSW mixing angles and effective eigenstate masses are changed by the postulated effect of a weak field in matter. The subscript notation is peculiar to this section. We point out that the constraint of mass conservation, or of mass-expectancy conservation, nowhere is imposed.

Step 1. We assume a theoretical expression such as (4.1.1) below used to represent creation of a neutrino with flavor $f \in \{e, \mu, \tau\}$ at space-time coordinates $\vec{0}$. So, a flavor state expression $\langle m_f(\vec{0};0) | m_{f'}(\vec{x};t) \rangle$ may be used to represent the transition amplitude $A(f \to f';t)$ to flavor $f'$ after time $t$. For mass states $m \in \{1,2,3\}$, using a delta functional $\langle v_m | v_{m'} \rangle = \delta_{mm'}$, we then may write the transition probability (oscillation intensity) $P$ in terms of transition matrix elements as,

$$P(f \to f';t) = |A(f \to f';t)|^2 = \left| \sum_m V_{f'm} V_{fm}^* \, m_m(\vec{0},0) m_m^*(\vec{x},t) \right|^2. \tag{2.8}$$

This is Eq. (7.26) of [5]. The mass amplitude vectors determine the flavor state transition.

Step 2. The initially-created, freely propagating neutrino is assumed well approximated as a plane wave. A wave-packet approach [5, Chapter 9; 6] similarly leads to a formula describing oscillations, but we choose the plane wave here for simplicity. For a mass eigenstate $m_m(\vec{x};t) \equiv \psi_m(\vec{x};t) | m_m \rangle$, $m = 1,2,3$, with initially well-defined energy and momentum, we have,

$$m_m(\vec{x};t) = \exp(i\vec{p}_m \cdot \vec{x} - iE_m t), \tag{2.9}$$

which is Eq. (7.27) in [5].

Step 3. Because neutrinos are ultrarelativistic, the momentum may be treated as though greater in magnitude than the mass, *viz.*, $|\vec{p}_m| \gg m_m$. This might seem to be an odd assumption from a kinematic point of view, because the units are disparate, and according to the Lorentz factor $\gamma$, mass scales up in importance as the velocity approaches the speed of light. However, this assumption is very reasonable in terms of a scattering process in which momentum supplies energy to be allocated to a population of particles of known, relatively small rest mass.

In either case, we digress here briefly to examine momentum and mass more closely: For a particle of fixed rest mass $m > 0$, the relativistic one-dimensional $p = (m g) u$. Therefore,

$$\frac{\partial p}{\partial m} = g u ; \text{ and,} \tag{2.10a}$$

$$\frac{\partial p}{\partial u} = m \frac{\partial}{\partial u}\left[ u\left(1 - \left(\frac{u}{c}\right)^2\right)^{-\frac{1}{2}} \right] = g^3 m. \tag{2.10b}$$

Forming the ratio,

$$\lim_{u \to c}\left( \frac{\partial p / \partial u}{\partial p / \partial m} \right) = \lim_{u \to c}\left( \frac{g^3 m}{g u} \right) = m \cdot \lim_{u \to c}\left( \frac{g^2}{u} \right) = \infty. \tag{2.10c}$$

So, as $u \to c$, a small change in $u$ gains a much greater effect on the momentum than a similar small change in $m$.

But, if $[x, p] = i\hbar$ and the uncertainty $\Delta x$ in distance of propagation is small, then neutrino flavor oscillations would imply that $\Delta u$ (determining the phase at observation distance $x$) also was small; therefore, large uncertainty in the momentum should be manifested in comparatively large mass uncertainties $\Delta m$.  We use this idea in Section 3, below.

Returning to the third step of the oscillation formula derivation, we accept without further analysis that $|\vec{p}_m| \gg m_m$; and so, calling $p_m = |\vec{p}_m|$, we proceed to Eq. (7.28) of [5],

$$E_m = \sqrt{p_m^2 + m_m^2} \cong p_m + \frac{m_m^2}{2 p_m}. \tag{2.11}$$

Step 4.  Taking the direction of propagation to be exactly on the $x$-axis, we may combine Eqs. (2.9) and (2.11) to write,

$$m_m(x;t) \cong \exp\left( i p_m (x - t) - i t \frac{m_m^2}{2 p_m} \right). \tag{2.12}$$

Step 5.  Using this last in (2.8) above,

$$P(f \to f';t) = \left| \sum_m V_{fm} * \exp\left( i p_m (x - t) - i t \frac{m_m^2}{2 p_m} \right) V_{f'm} \right|^2. \tag{2.13}$$

Step 6.  Assuming conservation of momentum, and letting $x \to t$ as $\mathbf{u} \to c$, we may square out (2.13) to obtain Eq. (7.32) in [5],

$$P(f \to f';t) = \sum_m |V_{fm}*|^2 |V_{f'm}|^2 + 2\operatorname{Re}\sum_{m \neq m'} V_{fm}* V_{f'm} V_{fm'} V_{f'm'} * \exp\left(-it\frac{\Delta m_{mm'}^2}{2p}\right), \quad (2.14)$$

in which the familiar abbreviation, $\Delta m_{mm'}^2 \equiv m_m^2 - m_{m'}^2$, first appears.

Step 7.  Again ignoring units, we may let $p \to E$ as $\mathbf{u} \to c$; we already have $t = x$. As in Eq. (7.40) of [5], assuming CP-violation negligible, we may ignore it in a simplification of $\vec{\vec{V}}$ to represent two-flavor mixing by angle $q$,

$$\vec{\vec{V}}*(q) = \vec{\vec{V}}(q) = \begin{bmatrix} \cos q & \sin q \\ -\sin q & \cos q \end{bmatrix}, \quad (2.15)$$

with $\det[\vec{\vec{V}}(q)] \equiv 1$.  Now, using (2.14) and (2.15) to represent oscillation between just two flavors, we get Eq. (7.40) in [5]:

$$P(e \to \mathbf{m}, x = L) = \frac{1}{2}\sin^2(2q) - \frac{1}{2}\sin^2(2q)\cos\left(L\frac{\Delta m_{21}^2}{2E}\right); \text{ or,} \quad (2.16)$$

$$P(e \to \mathbf{m}) = \sin^2(2q)\sin^2\left(L\frac{\Delta m_{21}^2}{4E}\right). \quad (2.17)$$

This result, the usual two-flavor oscillation formula, clearly will not be linear in $L$ so as to preserve transition-probability ratios.  The more complicated case of three-flavor oscillation is given by Eq. (7.46) in [5]; it includes three linear combinations of terms like the rightmost factor of (2.17) and similarly is nonlinear in distance.

### 2.2.1.2  Decay

We know that the ~15 MeV electron antineutrino burst from supernova SN1987A preceded the arrival of the first photons by less than a day over a lab-frame flight time of some 150,000 years [7].  With reasonable assumptions about supernova dynamics, the expected rest mass of the electron neutrino therefore likely would not exceed some 10 $eV/c^2$.  Astrophysical mass estimates lately have tended closer to 1 $eV/c^2$ [8].

The primary problem with postulating neutrino decay as a relevant mechanism for the upward-going $\mathbf{n_m}$ atmospheric deficit, is the presumed very small rest mass of the neutrino: There aren't many ways it could decay.  Of course, if the mass were zero, an observation instrument could not be moved quickly enough to measure both at the initial and final interaction points, observable events could not be ordered solely in time, no proper time would elapse, and the neutrino, like the photon, never would decay in vacuum.

The majority of oscillation theory variants require all three neutrino types to have mass eigenstate differences $|\Delta m_{mm'}|$ of fractions of an $eV/c^2$. Because the eigenstates are supposed to propagate as trios of free particles, conservation of momentum might seem to imply that each eigenstate have about 1/3 the expected mass of the propagating neutrino. Further reasoning along these lines points up the superposition vs. kinematic conflict theme of the present work but is deferred until the next Section.

Pending discovery of yet lighter fermions than neutrinos, we might as well assume that there is a decay of the neutrino mass eigenstates by a photon radiative mechanism. Problems with this are discussed in [5, Chapter 12]; in general, shoring up the theory to sidestep the low calculated rate makes the Standard Model or the neutrino population, or both, more complicated. The possible decay of neutrinos has been treated at length by Lindner, *et al* in [12], who also have recognized some of the oscillation kinematic issues discussed in Sections 3 and 4 below. Winter [19] also has calculated some interesting implications of neutrino decay.

So, for radiative decay, assuming an electron-rich medium and $m_3 > m_2 > m_1$, we might look at the decay process,

$$m_2(k_2) \to m_1(k_1) + g(q),  \tag{2.18}$$

in which the 4-momentum $k_2 = k_1 + q$. To explain the atmospheric deficit, we most simply would assume mixing such that $m_2$ dominates the mass vector when $f_m$ is detected. If so, then integration over the distance of the diameter of the Earth, compared with the integration over the nearby atmosphere, would show a nonlinear decrement in $P_{v_m}$, accounting for the observations.

The theoretical decay rate of a neutrino may be assumed higher in matter than in vacuum, but then the difference between decay and scattering becomes a matter of words. Specifically, even massless neutrinos may be treated as though "decaying" during passage through matter: Any interaction with the medium implies causality, which always adds a nonzero component to the overall proper time.

It is recognized that this statement opens issues not necessary to the rest of the presentation, but we merely mention here that interactions cause overall propagation to lag the light cone, thus allowing a synchronizing measuring instrument to be applied both at the initial and final interaction points. So, scattering makes events orderable purely in time and hence introduces a nonzero proper time of overall propagation. Depending on personal preference, one simply may insist that every scattering event of a massless particle is an ordered annihilation plus creation event, which avoids the dilemma of allowing a massless particle to propagate at a speed lower than *c*.

### 2.2.2  Angle (refractive or elastic scattering)

Let us treat the Earth as a neutrino lens. Letting neutrino refractive index increase with density, we may use upward angle $a_1 \cdot a_2$ and downward angle $b_1 \cdot b_2$ as defined in Section 2.1.2 above, to describe the detector input aperture (pupil). We then may treat the image of the far-side, upward, atmosphere as being magnified relative to the nearby, downward atmosphere. We may choose cones of observation along a radius of the Earth passing through Super-K, as in Fig. 2 above. Then, calling the downward atmosphere as the initial field, and upward as final field, we may define a transverse magnification factor $m_T$ as,

$$m_T = dx_f/dx_i = dy_f/dy_i .\qquad(2.19)$$

Assuming conservation of étendue as in [4, I.1.7, Eqs. (71) and (72)], if the initial and final planes are taken as though conjugate with transverse magnification $m_T$, we have,

$$n_i^2 d\mathbf{b}_1 d\mathbf{b}_2 = n_f^2 m_T^2 d\mathbf{a}_1 d\mathbf{a}_2; \text{ and, so,}\qquad(2.20)$$

$$\frac{\int_{\substack{\text{entrance}\\ \text{pupil}}} d\mathbf{a}_1 d\mathbf{a}_2}{\int_{\substack{\text{exit}\\ \text{pupil}}} d\mathbf{b}_1 d\mathbf{b}_2} = n_{fi}^2 m_T^2,\qquad(2.21)$$

in which $n_{fi}$ is the ratio of final to initial refractive index.

The total power $\Phi$ in an optical system from a source is given [4, I.1.88, Eq. (303)] as,

$$\Phi = \frac{1}{p} m_a \int_{\text{field}} dxdy \int_{\text{pupil}} d\mathbf{a}_1 d\mathbf{a}_2 .\qquad(2.22)$$

If we look at the power ratio of two distinct sources of species 1 and 2, superposed and of identical size and location, using the same optical system, the observation angle integrals in Eq. (2.22) drop out as in Section 2 above, as do the field integrals within the magnification factor, and we get a relation analogous to Eq. (2.4) above,

$$\Phi_{\text{bias}} = \frac{\Phi_1}{\Phi_2} = n_{fi}^2 m_T^2 ,\qquad(2.23)$$

and any asymmetry in the flux may be expressed by the product of refractive index and magnification for the two species.

So, we have two different but not independent factors to explain the observed $\mathbf{n}_m$ flux deficit: (*a*) <u>refraction</u> by the bulk of the Earth, a property intrinsic to the matter of which it is composed; and, (*b*) <u>magnification</u>, a geometrical property of the spherical shape of the Earth. In thermodynamics, there is a similar distinction made between intensive and extensive variables. We mention that scattering equations developed for neutrino-(*other particle*) interactions can shed light only on the refractive-index term in (2.23) above; geometry has to be accounted for separately.

We also note that the two index-and-shape factors might combine near the horizon, causing a zone of total internal reflection which would increase the $v_m$ flux above the expected 2:1 ratio versus the $n_e$ flux.

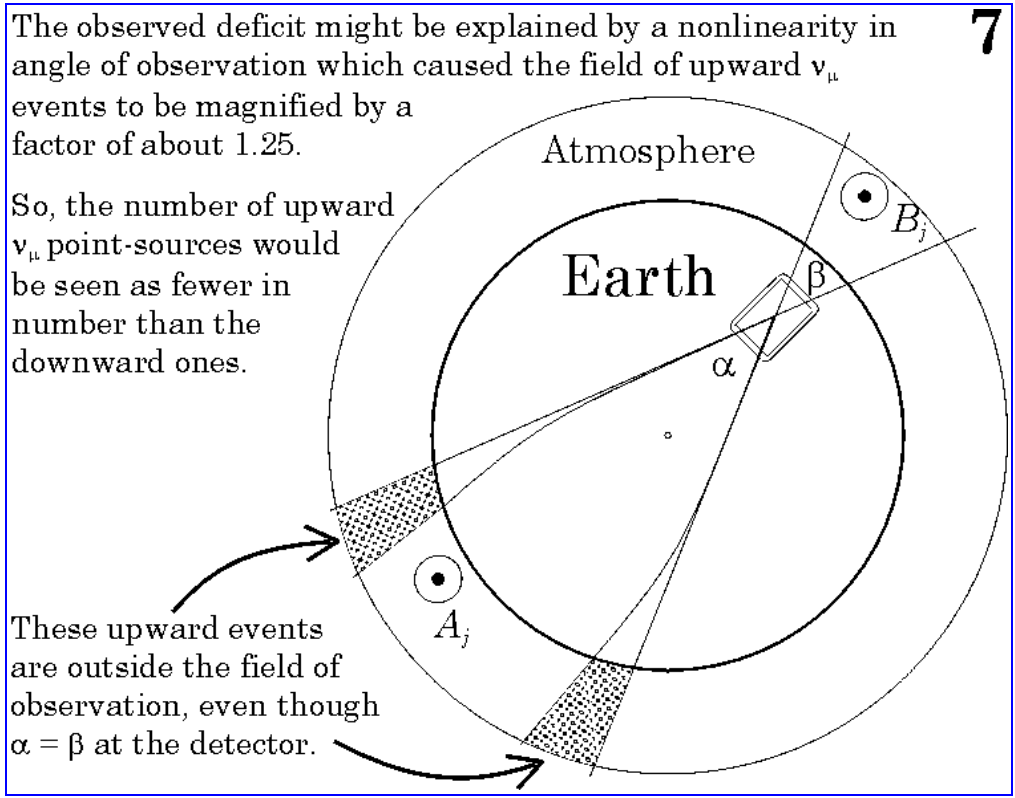

**7**

The observed deficit might be explained by a nonlinearity in angle of observation which caused the field of upward $v_\mu$ events to be magnified by a factor of about 1.25.

So, the number of upward $v_\mu$ point-sources would be seen as fewer in number than the downward ones.

These upward events are outside the field of observation, even though $\alpha = \beta$ at the detector.

But, to account for the observed flux deficit in observation directions not near the horizon, we would need some new physics: For neutrinos of several GeV, the observed deficit means the field of view would include only about 2/3 the number of point-source $n_m$ events; so, $n_{fi}m_T$ in Eq. (2.23) above would amount to about 0.8, implying an effective linear magnification or refractive index (or their product) of about 1.25.

This would seem to be a huge interaction for neutrinos, at least for the more studied electron neutrinos. Recall that the range of the weak force has been quoted [10, p. 123] at about $r_{weak} = 10^{-17}$ m. So, the volume $V$ of a cylinder with radius equal to the weak range and with length the diameter of the Earth would be,

$$V_{weak} = p \cdot r_{weak}^2 \cdot 2R = p \cdot 10^{-17 \cdot 2} \cdot 12.8 \cdot 10^6 \cong 4 \cdot 10^{-27} \text{ m}^3. \qquad (2.24)$$

Given an average Earth density of $5.5 \cdot 10^3$ kg/m$^3$, each upward muon neutrino would have passed in weak-range only of some $10^3$ to $10^4$ nucleons. Either the neutral-current cross-section would have to be far greater than expected, or the electronic interaction would have to be far stronger than expected, somehow. The MSW hypothesis also postulates a stronger charged-current interaction than might be expected from scattering data under $W^\pm$ exchange.

# 3. Neutrino Oscillation Observations

## *3.1 Source - Detector Distances*

### 3.1.1 Oscillation phase must be well defined

If just momentum or energy were conserved, we could assume one of them constant, and let **u** (or **u**$^2$), as calculated under the usual theory, vary inversely with mass expectancy. But both are conserved for a free particle, which leads to the $v$ and **u**$^2$ paradox (below). So, for a free particle, if $m$ varies between initial and final neutrino interaction, then we should assume $v$ doesn't. If neutrino types are assumed to differ in mass, this assumption is the only way to guarantee that the oscillation phase might be well defined, because with $v$ not varying much, the mass expectancy could change accurately according to theoretical (*viz.*, calculated) distance.

Along these lines of reasoning, the more relativistic the neutrino, the more the mass determines the kinematics, because of the Lorentz scaling of mass (but not of velocity). Of course, where $v$ is allowed to vary, the $v$ dominates the kinematics, as shown in the notes on oscillations, above.

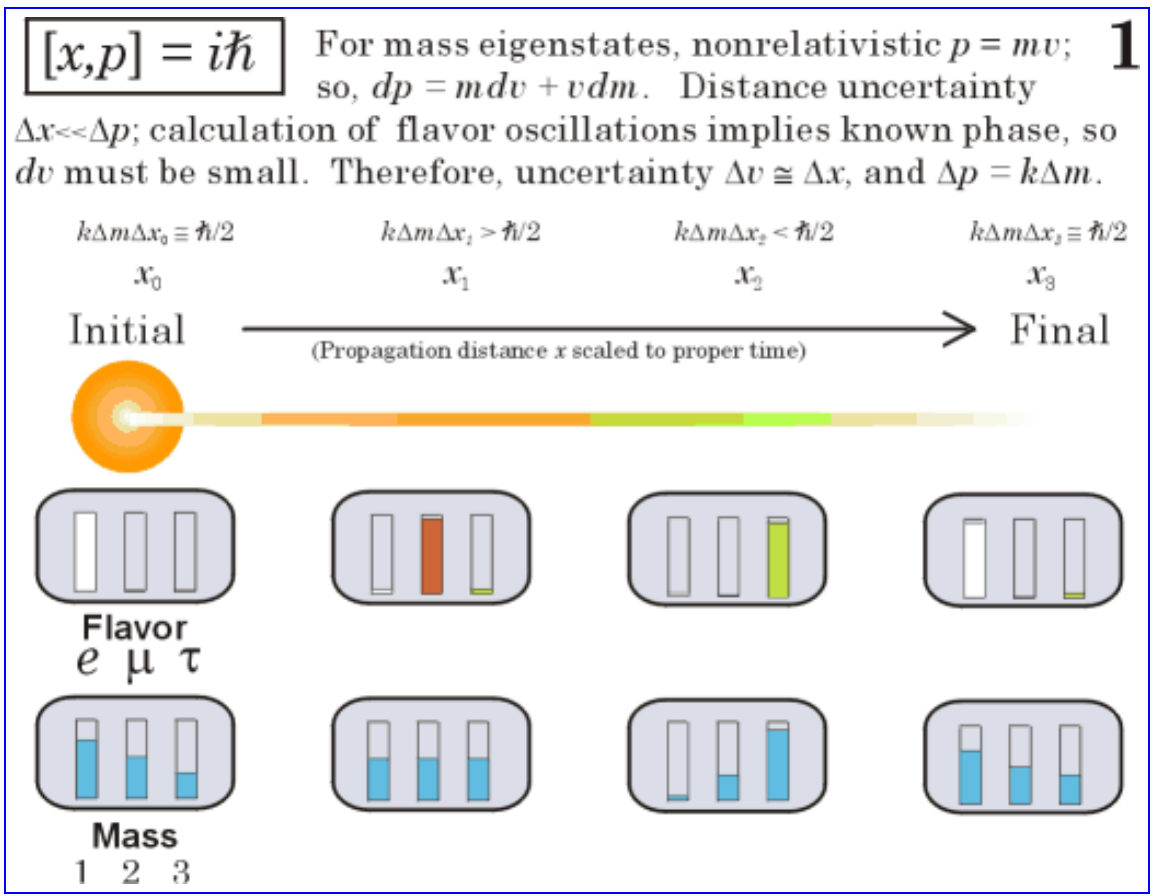

# 4. Vacuum Oscillation Theory from CKM

After looking at a few common expressions of the vacuum flavor-mixing theory, we present two separate problems with the result: (1) The problem of the short range of superposition the weak force; and, (2) the problem of conservation of mass of a freely propagating neutrino.

## *4.1 The Usual Expression*

As in [5], Chapter 7, ignoring time-evolution and viewing the problem strictly in terms of distance $x$ from the neutrino initial (creation) point, we write the flavor state as a function of detector distance $x$ in terms of the mixing matrix $\ddot{\mathbf{V}}$ and the mass state (mass-eigenstate state vector):

$$\vec{f}(x) = \ddot{\mathbf{V}} \cdot \vec{m}(x), \text{ with } m_j(x) \equiv \mathbf{y}_j(x) \big| m_j \rangle; \text{ or,} \tag{4.1.1}$$

$$\begin{pmatrix} f_e \\ f_m \\ f_t \end{pmatrix}(x) = \begin{bmatrix} V_{e1} & V_{e2} & V_{e3} \\ V_{m1} & V_{m2} & V_{m3} \\ V_{t1} & V_{t2} & V_{t3} \end{bmatrix} \cdot \begin{pmatrix} m_1 \\ m_2 \\ m_3 \end{pmatrix}(x) = \begin{bmatrix} V_{e1} & V_{e2} & V_{e3} \\ V_{m1} & V_{m2} & V_{m3} \\ V_{t1} & V_{t2} & V_{t3} \end{bmatrix} \cdot \begin{pmatrix} \mathbf{y}_1(x)\big| m_1 \rangle \\ \mathbf{y}_2(x)\big| m_2 \rangle \\ \mathbf{y}_3(x)\big| m_3 \rangle \end{pmatrix}. \tag{4.1.2}$$

We assume exactly three flavors of neutrino. We note that the mass state vector, $\vec{m}(x)$, as a function of distance, is what is supposed to cause observation of the atmospheric muon neutrino deficit by a change in flavor expectancy according to formula (4.1.1). Formulas for mixing as a function of distance usually are derived in terms of the mass eigenstate squared-mass difference, $\Delta m_{ij}^2 = \big| m_i^2 - m_j^2 \big| = \big| \langle m_i(x) | m_i(x) \rangle - \langle m_j(x) | m_j(x) \rangle \big|$. The state vector components are allowed to vary in phase, which permits each one to have constant mass during propagation. See Section 2 above for more details on the formula itself. We note here that because the variance of a difference of random variables is equal to the variance of the sum, the Heisenberg uncertainty involved in a mass difference is not different in any essential way from the uncertainty in the mass state $\vec{m}(x)$ itself.

A few examples of (4.1.2):

For no mixing and therefore no flavor oscillation,

$$\begin{pmatrix} f_e \\ f_m \\ f_t \end{pmatrix}(x) = \begin{bmatrix} V_{e1} & 0 & 0 \\ 0 & V_{m2} & 0 \\ 0 & 0 & V_{t3} \end{bmatrix} \cdot \begin{pmatrix} m_1 \\ m_2 \\ m_3 \end{pmatrix}(x); \text{ or, equivalently,} \tag{4.1.3}$$

$$\begin{pmatrix} f_e \\ f_m \\ f_t \end{pmatrix}(x) = \begin{bmatrix} 1 & 0 & 0 \\ 0 & 1 & 0 \\ 0 & 0 & 1 \end{bmatrix} \cdot \begin{pmatrix} m_1 \\ m_2 \\ m_3 \end{pmatrix}(x) \implies \big| \vec{f} \rangle = \big| \vec{m} \rangle. \tag{4.1.4}$$

For maximal mixing [5, Eq. (7.50)],

$$\vec{f}(x) = \begin{bmatrix} 1 & -1 & -1 \\ 1 & e^{i2p/3} & e^{i4p/3} \\ 1 & e^{i4p/3} & e^{i2p/3} \end{bmatrix} \cdot \vec{m}(x), \qquad (4.1.5)$$

noticing that the determinant of this matrix is $i\sqrt{3}$.

For an exact analogy to Cabibbo-Kobayashi-Maskawa (CKM) quark mixing [9, (2.16)],

$$\vec{f}(x) = \begin{bmatrix} 0.98 & 0.22 & 0.003 \\ 0.22 & 0.97 & 0.04 \\ 0.008 & 0.04 & 0.99 \end{bmatrix} \cdot \vec{m}(x), \qquad (4.1.6)$$

which has determinant $\cong 0.89$.

For a large mixing angle solution to our atmospheric problem, [18, (6)],

$$\vec{f}(x) = \begin{bmatrix} 0.825 & 0.565 & 0 \\ -0.400 & 0.583 & 0.707 \\ 0.583 & -0.400 & 0.707 \end{bmatrix} \cdot \vec{m}(x), \qquad (4.1.7)$$

which has determinant $\cong 0.97$.

## *4.2 Superposition vs. Interaction*

Simple, coherent, Young-type superposition has to be stretched to be applied to neutrinos, which are supposed to propagate as trios of mass eigenstates not confined to short distances. This difficulty is because of the small source aperture relative to propagation distance.

### 4.2.1 Photon Waves

> **Young-type Superposition of Amplitudes:**   **2**
>
> Massless Heisenberg: $4\pi \Delta p \Delta x \geq h \;\Rightarrow\; 4\pi \Delta(h\nu)\Delta x \geq h \;\Rightarrow\; 4\pi \Delta x \geq \Delta\lambda$.
>
> [diagram: lamp at left with slit spacing $\Delta x_i$, distance $r$ to screen at right, angle $\theta$, $r_\theta = \sqrt{r^2 + (\Delta x_i)^2}$, with wave showing $\Delta x_f = \Delta\lambda$]
>
> Let $r_\theta - r \equiv \Delta r = \sqrt{r^2 + \Delta x_i^2} - r$. Then, $r = \Delta x_i^2/2\Delta r - \Delta r/2$. The second term will be negligible, so $r \cong \Delta x_i^2/2\Delta r$, and,
>
> $\Delta r \equiv$ Heisenberg $\Delta x \geq$ (*some limit*) $\;\Rightarrow\; r <$ (*that limit*).
>
> Call $\Delta\lambda = 10^{-7}$ m for visible light. Then $\Delta r \geq 10^{-6}$ m; so, for slit spacing $\Delta x_i = 10^{-3}$ m, we get approximately $r < 50$ m. For a $E = 1$ GeV neutrino with classical Compton wavelength $hc/E$, $\Delta r \cong 10^{-17}$ m, making $r < 5 \cdot 10^{10}$ m for the same slit spacing.

At a reasonably long distance $r$, plane wave and wave-packet representations of a particle wavefunction make no difference to the calculations in Poster #2. Recall that for a massless particle, the quantum $p = h\mathbf{n}/c = h/(c\mathbf{l})$. Then, $E = pc = h/\mathbf{l}$. All the photons in Poster #2 above are assumed coherent, so perhaps the lamp would represent a laser. The slit spacing of 1 mm in Poster #2 would be huge for a process rendered coherent solely by the weak force.

### 4.2.2 Neutrino Waves

If neutrinos, or, in the usual oscillation theory, neutrino mass eigenstates, were massless, then the same calculations would hold as for photons. If they were massive but not very much so (*m* very small), then a massless calculation might be used as an approximation. This is really what the previous discussion of "$|p| \gg m$" is about.

> **Oscillatory Superposition of Eigenstates:** **3**
>
> [Diagram showing $\Delta x_i$ on left, $r$ across top to screen on right with $\Delta x_f$, angle $\theta$, and $r_\theta = \sqrt{r^2 + (\Delta x_i)^2}$]
>
> If $r \cong \Delta x_i^2 / 2\Delta r$ with $\Delta x_i \cong 10^{-17}$ m (range of the weak force), from the Compton $\Delta\lambda$ of a 1 GeV neutrino, we would get $r < 5 \cdot 10^{-18}$ m.
>
> For a massive free particle with a known velocity, Heisenberg gives $4\pi\Delta p \Delta x \geq h \Rightarrow 4\pi\Delta(mv)(\Delta x) \geq h \Rightarrow \Delta r \equiv \Delta x \geq h/(4\pi v \Delta m)$. So, if $v \to c^-$, we get $r < (\Delta x_i^2 \cdot 4\pi c^- \Delta m)/(2h) \Rightarrow r < c^- \Delta m$; and, $r < \sim 5 \cdot 10^{-11} \Delta m$ m, for $\Delta m$, in eV/$c^2$, the approximate uncertainty in the invariant mass.
>
> Relativistically, $v < c$; but, approximating with $r < 5 \cdot 10^{-11} \gamma \Delta m$ m, we may take $r$ near the $12 \cdot 10^6$ m diameter of the Earth to find $\gamma \cong 10^{17}$ $\Rightarrow v \cong c\sqrt{1 - 10^{-34}} \Rightarrow \Delta m^2 \cong 10^{-16}$ (eV/$c^2$)$^2$, a 1 GeV-neutrino rest mass difference (squared) that seems low for oscillations. What is wrong?

    So, we end up with a problem because of the short range [10, p. 123] that the weak force is believed to have. If the hypothetical three neutrino mass eigenstates propagated as though a single particle, rather than as a field superposition, this problem would not arise, because we could allow the field to be unbound by the interaction, and to superpose its elementary components very close to the final interaction point.

    Note that the uncertainty in Poster #3 applies to any measurement of two relativistic or massless particles, even if created at a single point and with identical momenta. Two or more 1-GeV relativistic particle wavefunctions created by a force not extending over a range greater than $10^{-17}$ m can not be used to explain the atmospheric problem, unless the particles were far less massive than anything which might be fit by the usual oscillation theory.

    Some confusion is possible here: The range of the weak force as a field is assumed independent of the hypothetical mass-eigenstate wavepacket size or of the actual occurrence of a *W* or *Z* particle exchange.

    Consider the Young's experiment as above: The relative location (phase) of the propagating mass eigenstates is analogous to the distance between slits and here is applied to transverse displacement. The superposition of mass eigenstate amplitudes represented by the $\vec{\tilde{V}}$ matrix input vector wavefunctions (Eq. 4.1.1) is analogous to the superposition of photon amplitudes at some specific location on the screen; a photon approaching the slits is assumed to have a field (not determined by a virtual photon exchange) which allows it to superpose amplitudes through both slits, creating an interference, as opposed to a one-slit diffraction pattern.

That this is the correct analysis in quantum mechanics has been shown for two-atom interference patterns in [15]; so, it hardly can fail to be correct for superposition of amplitudes of propagating mass eigenstates, too.  The weak force is supposed to have a finite range ($\cong m_W^{-2}$; certainly below $10^{-17}$ m), unlike the infinite range of the electromagnetic force; beyond this range, something analogous to diffraction, but not interference among mass eigenstates, is possible in the usual neutrino oscillation theory. So, there are two new limitations when comparing neutrinos with photons: (*a*) At creation, the initial aperture is limited to one spannable by the range of the weak force; and, (*b*) during propagation and at final interaction, the superposition is limited by the range of the weak force.  The mass eigenstates in the usual theory interact at a final location determined by their position wavefunctions; however, the weak force must determine the mixing superposition (if any), not the particle wavefunction(s).

In contrast to interference by quantum superposition, an exchange of *Z*, say, would indicate weak scattering, which we do not require for the superposition postulated in the usual oscillation theory.  After propagation for a long distance, by the analogy with Young's experiment, we assume the neutrino might "encounter" a set of mass eigenstates separated too far for them to superpose weak fields.  Such a neutrino would not show oscillation--and, really, it shouldn't be allowed to show up at all!

### 4.2.3  Neutrino Long-Distance Eigenstate Separation

This limiting case of decoherence with distance of propagation, which possibly better would be called *separation* of the mass eigenstate position wavefunctions, is an interesting implication of the usual theory, provided the reasonable assumption be made of a wave-packet representation.  Of itself, this separation is not inconsistent with other physics, but only if interpreted carefully.  Most importantly, there can be no multiple final interaction, one per mass eigenstate, for a long-range, oscillating neutrino.  But, there can be no entangled collapse of wavefunctions, either, because if the mass eigenstates superpose in an evolving process, by definition they can not be entangled. Instead, consider the details of what would have to be the final interaction:

First, ignoring gravity as usual, the final interaction has to be mediated by the weak force.  Second, if we are to accept the oscillation theory, there must be a superposition of mass eigenstates at the point of final interaction.

A (position) wavefunction represents just one thing, the amplitude of finding the particle at a given point in an interaction.  So, for any oscillation-theory neutrino to interact, it must interact in a volume of position space no larger than that given by the range of the weak force, and limited by the joint probability that all three mass eigenstates will be found in that volume.  Furthermore, it will interact with its target particle (say, an electron) only if the amplitude of that particle's position wavefunction overlaps in that same tiny volume.  This represents the very small cross-section expected of a neutrino, and a cross-section which we now see necessarily must decrease with propagation distance, as the individual mass eigenstate wave packets gradually separate and overlap less and less.

The end point of a forever-propagating neutrino, then, according to the oscillation theory, would be three widely-separated mass eigenstate wave packets which overlapped so little that the probability of superposition at the same location as a target particle would be as good as 0. So, we would end up not with a sterile neutrino (sterile by flavor), but with a *stale* neutrino, because of mass eigenstate dispersion.

## *4.3 Free Propagation*

Requirements are:

- nonvirtuality (eigenstates must be real particles, not virtual).
- binding interaction, for nonelementary particles (ignored in the usual neutrino oscillation theory).
- conservation of energy, momentum, and, therefore, mass.

Contrast the quark sector, in which the particles never are free. Quarks always are bound in a local field by gluon exchange (strong force); quarks also interact by photon exchange (coulomb force).

### 4.3.1 The $v$ and $v^2$ Paradox

> **4**
>
> A freely propagating relativistic particle obeys $E^2 = (pc)^2 + (mc^2)^2$.
>
> Energy is conserved, so, $(k_{energy})^2 = (mc)^2((v\gamma)^2 + c^2) = m^2((v\gamma)^2 + c^2)c^2$.
>
> Momentum is conserved, so, $(k_{momentum})^2 = (mv\gamma)^2 = m^2(v\gamma)^2$.
>
> Therefore, because virtuality is not allowed for free particles, mass is conserved, and no observation is possible showing an effect depending on a change in mass expectancy as a function of free propagation distance.
>
> For oscillation, we must invoke the Heisenberg uncertainty principle to allow the mass-vector expectancy to change during flight; only this way can we perform a CKM-like computation of a change in flavor expectancy as a function of distance in vacuum.
>
> The uncertainty is not in the eigenvector weight amplitude, but in the mass; so, in the previous, $k \equiv 1$ and $\Delta m \Delta x_j \to \hbar/2$.

Some confusion is possible here, also. The **y** factors in (4.1.1) above determine relative mass phases. Each mass eigenstate, according to the usual theory, has a constant mass which is weighted by the amplitude of the respective **y**, like the amplitude of a photon through a slit in Young's experiment. The intensity of light, or, by analogy, the flavor-determining mass of the neutrino, is given by the vectorially summed amplitudes at the final point of observation (neutrino annihilation or scattering). This sum can not be constant during propagation (in range of the weak force) any more than could be the intensity of light in an interference pattern.

The constant mass of each mass eigenstate is analogous to the constant intensity through each slit, as summed and measured in a region close to the slit (or calculated from the area of the slit). Therefore, by the superposition hypothesis of the usual oscillation theory, the mass amplitudes will have to change as a function of distance of propagation so that their real parts always remain constant (otherwise, they would not be eigenstates) and also so the real part of the mixed neutrino mass remains constant (otherwise, mass would not be conserved). This means that only the phase of the state vector components may change during propagation.

The usual oscillation theory requires a change in the real part of the flavor vector, to change the observed flavor expectancy as a function of neutrino propagation distance. This may be seen easily from (4.1.1) above:

$$\vec{f}(x) = \vec{\vec{V}} \cdot \vec{m}(x). \tag{4.3.1}$$

The mixing matrix usually is assumed unitary; however if it merely is nonsingular and invertible, we may clarify the algebra by solving formally for the mass vector:

$$\vec{\vec{V}}^{-1} \cdot \vec{f}(x) = \vec{\vec{V}}^{-1}\vec{\vec{V}} \cdot \vec{m}(x); \text{ so,} \tag{4.3.2}$$

$$\vec{m}(x) = \vec{\vec{V}}^{-1} \cdot \vec{f}(x). \tag{4.3.3}$$

And, according to the usual theory, the value of flavor is observed to change as a function of distance, while the calculated value of the mass of the neutrino may not change during propagation.

This in turn means either that the mass eigenstates all must be equal in mass, or that the mass of the neutrino must not be a function of flavor. The usual theory requires different masses for the mass eigenstates; so, it appears that it must be, by the usual theory, that <u>all neutrino flavors have equal mass expectancy</u> (and possibly equal mass).

This is an inconsistency which has not been addressed adequately so far in the literature. We argue by analogy to quarks or the signed leptons that there may be a neutrino mass hierarchy; yet, we reject that mass hierarchy to make the superposition matrix work. And, superposition doesn't work because of the Heisenberg principle as discussed in Section 4.2 above.

There is mass hierarchy, if not simple mass difference, everywhere except in the neutrino sector. However, if we allow for mass differences in oscillating neutrinos, then the mass in the final interaction will differ from that in the initial interaction, leading back to the *u* and *u*$^2$ paradox. The revised theory below tries to sidestep this problem.

### 4.3.2 Uncertainty Should Depend only on the Commutator

> But, it's worse! We need A to permit oscillation; however, oscillation to a calculated flavor implies B.
>
> **5**
>
> Energy uncertainty in final-state neutrino, from [x,p]
>
> ⟵———A
>
> $\langle m_i \rangle$     $\langle m_f \rangle$
>
> (Excluded, by energy conservation at $\langle m_f \rangle$ during oscillation)
>
> Energy uncertainty from $mv^2$ and [x,p]
>
> ⟵———B
>
> $\langle m_f \rangle$
>
> So, we have a contradiction; and, CKM-like oscillations can not occur for free particles.

To allow different masses for different neutrino flavors, and to address the "***u*** and ***u***$^2$ paradox" above, a common idea is to rank the importance of energy and momentum conservation to see which one to assign with certainty. For example, giving precedence to energy, one may <u>postulate</u> conservation of energy to calculate the mass eigenstate oscillation kinematics. Then, to sidestep the issue of the simultaneous conservation of momentum, which would imply the mass expectancy was the same in the initial and final states of each neutrino, one allows the uncertainty implied by $[x, p] = i\hbar$ to span the calculated neutrino mass difference. Note that this is a calculated difference, as *ca.* (4.1.2) above, not an observation itself subject to the Heisenberg principle. Also note that the uncertainty in the momentum implicitly is being assigned to the mass, not the velocity factor, which is consistent with the oscillation theory discussion above.

However, this idea is not tenable, either logically or physically, although the problem might not be immediately obvious. Here is the new problem:

Logically: If the interval of uncertainty spans both the initial and final calculated mass expectancies, how can it weight more heavily the interval of the initial expectancy during neutrino creation, and then more heavily the interval of the final expectancy during neutrino annihilation? Should not the observed flavor (which is determined by the mass vector, as explained above) vary randomly within the interval of uncertainty -- whether as a function of distance of propagation or as a function of anything else?

Physically: The claim under the energy conservation postulate is that because of the Heisenberg uncertainty principle, momentum is uncertain enough during neutrino (eigenstate) propagation that the interval of mass uncertainty spans at least the interval including the calculated initial and final neutrino mass expectancies. The oscillation theory calculations require that the interval of uncertainty change as a function of distance of propagation, so that the flavor might imply a different mass expectancy at different distances. However, the Heisenberg interval is a fundamental quantity and by definition never may be reduced below the commutator value. Therefore, allowing Heisenberg's interval to be a function of distance violates a fundamental law of quantum mechanics.

The argument just given, based on precedence of conservation of energy, may be repeated with identical results by beginning from a postulation of conservation of momentum and arguing from $[t, E]$. Neither the logical nor the physical problem can be avoided in the usual neutrino oscillation theory.

### 4.3.3 Breakdowns and Patchup Attempts by Others

Tsukerman raised the issue of the uncertainty conundrum immediately above in a criticism [11] of a paper by Guinti and Kim. In a later paper, Giunti [16] recognized and discussed the issues of "equal momentum" and "equal-energy" at length, rejecting both as meaningless (as previously had Zralek [20]) in terms of the derivation of the usual neutrino oscillation formula. However, Giunti did not consider the $u$ and $u^2$ paradox, which occurs for *all* energies and momenta, and proposed a new precedence of Lorentz invariance of "oscillation probability", based on the above Heisenberg reasoning, but simultaneously and coherently invoked both for momentum and energy.

Likewise, DeLeo, *et al* [13] have pointed out that neither momentum nor energy can be given precedence over the other and suggested that the only consistent solution would be to assume equal mass-eigenstate velocities. This doesn't work except for equal-mass neutrinos, either; however, it is consistent with the *revised neutrino* hypothesis presented below.

We also mention aside that many (e. g., [20]) have considered neutral kaon oscillations as the model on which to base neutrino oscillations as an explanation of the atmospheric and solar neutrino deficits. Whether or not the present argument might be sustained for kaons, one should notice that kaons decay while propagating, whereas neutrinos are assumed not to decay in the usual oscillation theory. See 2.2.1.2 above for more discussion of decay. Kaons therefore do provide a mechanism by which energy and momentum might be transferred independent of mass to the environment during propagation, thus invalidating the immediate applicability of the $u$ and $u^2$ paradox to their mass expectancies. Also, a very unusual effect, *CP* violation, is observed in conjunction with the kaon phenomenon; therefore, it is not clear that kaons might not be very exceptional both in *CP* conservation and in oscillation. Finally, as DeLeo, *et al* [13] point out, kaons do not propagate ultrarelativistically.

So, why copy kaons? The neutrino should not be assumed somehow analogous to the kaon on the issue of flavor oscillation. An oscillation theory of neutrinos must be justified on its own merits. The question of the coherence range raised in 4.2.2 above should be examined seriously by anyone subscribing to the oscillation hypothesis for kaons, but this issue will not be addressed further here.

# 5. Revised Vacuum Oscillation Theory

## *5.1 Based on Proper-Time Energy Progression*

- Neutrinos have energy substructure
- Neutrinos are massive
- Neutrino oscillations involve energy eigenstates
- If mass eigenstates exist, they must be virtual

## *5.2 Gives Substructure to Neutrino Interactions*

One way to avoid the **u** and **u**$^2$ paradox above, and the apparent problems with simple mass-eigenstate superposition, would be by substitution of energy eigenfunctions for mass eigenfunctions. The substitution might be called a ***revised theory*** of neutrino vacuum oscillations. In such a revision, the three neutrino types might be allowed to have distinct, definite masses $m_i$ as well as distinct flavors $f_j$; of course the mass expectancies $\langle m_i \rangle$ would be definite. In any case, each neutrino would have a specific mass determined entirely by its initial (creation) interaction. We note here that this is not a local reality theory, because we assume quantum-mechanical amplitudes to determine the interactions.

Here is how it might work:

Note: From here on, *i* and *f* subscripts refer to initial *vs*. final states, respectively.

A. All neutrino interactions would be assumed to occur in two stages. Both stages would be located within an interval small enough to contain the usual three mass eigenstates as virtual particles: The first stage we call the *flavor set*, and the second the *mass vertex*.

The flavor set always would precede the mass vertex, for neutrinos or antineutrinos, assuming these particles differed otherwise. The complete process is illustrated in the next poster sketch. No attempt is made here to analyze or diagram a Feynman-like propagator describing this process.

B. When a neutrino was created in its initial interaction, in the *flavor set*, it would be formed with the same three "mass eigenstates" of the current oscillation theory. The initial mass state would determine the initial flavor, as in Section 4 above; also, the revised neutrino would be created with its own, specific initial rest mass, $m_i$. Thus, at the end of the initial flavor set, a revised neutrino would have a definite flavor and a specific mass. It also would have a transient set of mass eigenstates. Its energy and momentum would be undefined.

C. Then, still during the initial interaction, a *mass vertex* would occur in which the revised neutrino's three mass eigenstates would be converted to weak energy states. A specific initial mass equal to $m_i$ would remain. The revised neutrino's weak field would create three energy eigenstates proportional to the respective energies of the three former mass eigenstates; these energy eigenstates might be called, a little facetiously, the three *quirks*. Radiative corrections might be applied, and the revised neutrino's quirk state, energy, momentum, and mass then would be determined, and this is how it would propagate as a free particle.

   D. During propagation, the revised neutrino's internal propagation dynamics (unitarity of $\vec{V}$) would cause the weak energy to be distributed on the three quirks exactly in the same amplitude that the mass phase was distributed among eigenstates in the current neutrino oscillation theory. In other words, as a function of proper time, according to Eq. (4.1.1) above, the as-yet unobserved flavor state amplitudes of the propagating revised neutrino also would change exactly as hypothesized in the current vacuum-oscillation theory. However, because the quirks are not massive, the neutrino's expected mass value $\langle m_i \rangle \equiv m_i = \sqrt{\langle m_i | m_i \rangle}$ would not vary as a function of distance.

   The neutrino's energy, momentum, and velocity would be measured as constant in expected value at any distance, as determined by its creation mass vertex. However, the fraction of $mc^2$ energy in the mass would be allowed to change from initial mass vertex $m_i$ to final mass vertex $m_f$, $m_i \neq m_f$, as explained below.

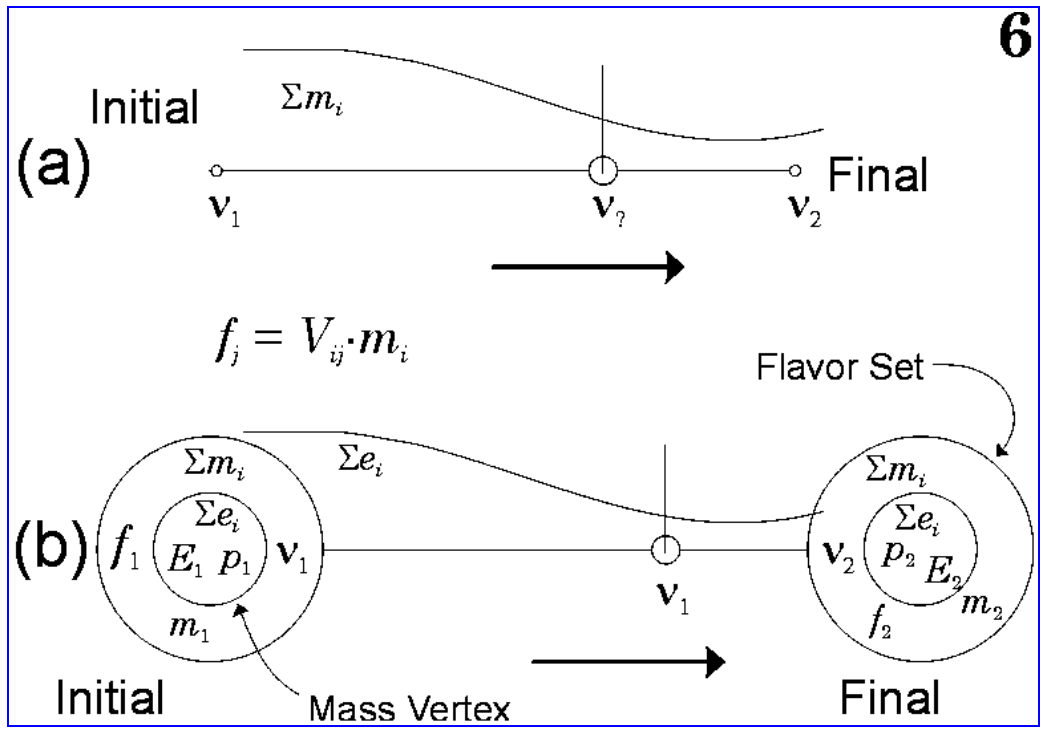

**Propagation of a neutrino created type 1, showing oscillation to type 2 at destruction.** (*a*) **Sketch of the current neutrino vacuum oscillation theory: The mass eigenstates $m_i$ superpose to determine the amplitude of flavor.** (*b*) **A revised neutrino theory which guarantees conservation of mass during propagation. The outer region must be entered first and determines flavor and mass, defining the type; kinematics then are determined from the inner region, which determines momentum $p$ and total energy $E$, as well as initializing the energy eigenstates.**

E. In its final interaction, the revised neutrino first would enter a *flavor set*: Here, flavor would be determined by the final flavor state as in Eq. (4.1.1) above, but with the quirks replacing the mass eigenstates. The weak cross-section would be given by the flavor expectancy as a function of propagation distance, as in the usual vacuum-oscillation theory. The propagation mass $m_i$ would become available virtually as energy, and some of the weak energy which quantized the quirks would be converted to the mass $m_f$ of the neutrino of final flavor. This would hold whether or not neutrinos of different types differed in mass, or had indefinite (superposed; mixed) mass. A set of transient mass eigenstates would be created (for compatibility with the current theory). We again would have a neutrino of a definite flavor and specific mass, but with undefined energy and momentum.

F. In the *mass vertex* of the final interaction, the revised neutrino's three quirks again would be formed from the mass eigenstates, and the momentum of the propagating neutrino would be preserved by a transfer of energy on the new final mass $m_f$ in the weak field. Thus, the final-state interacting neutrino would have the same momentum as in the initial state, and the same total energy as when it was created in the initial state. Its mass expectancy would be that of a neutrino of the final-state flavor (perhaps mixed), and observation would show scattering of a neutrino particle of this final mass $m_f$.

So, the final-state momentum, energy, and mass values would determine the kinematics and radiative corrections of the observed final interaction, which might be, for example, an (anti)proton interaction triggering ultimately a Cerenkov light cone in a neutrino detector.

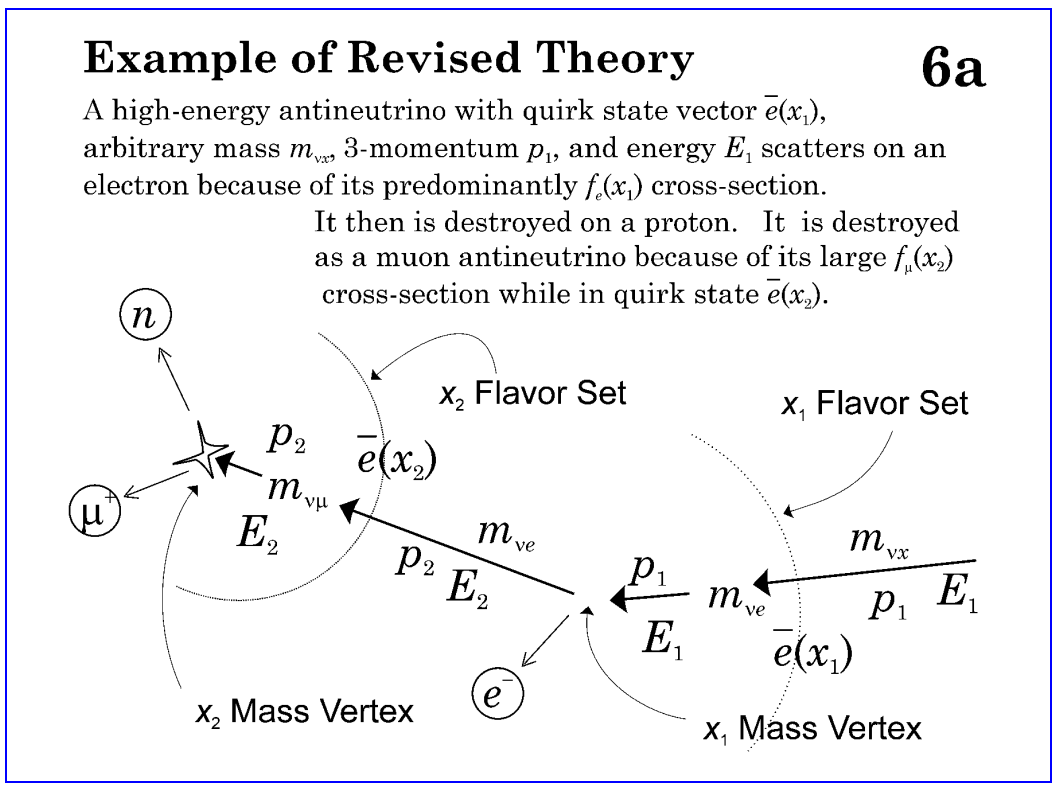

**Example of Revised Theory**  **6a**

A high-energy antineutrino with quirk state vector $\bar{e}(x_1)$, arbitrary mass $m_{vx}$, 3-momentum $p_1$, and energy $E_1$ scatters on an electron because of its predominantly $f_e(x_1)$ cross-section.

It then is destroyed on a proton. It is destroyed as a muon antineutrino because of its large $f_\mu(x_2)$ cross-section while in quirk state $\bar{e}(x_2)$.

We avoid here the question of whether neutrinos would appear transversely to be point particles during neutrino-neutrino or neutrino-(*other particle*) scattering. The revised theory implies that they would show some longitudinal extent in the direction of the momentum propagator, insofar as the Heisenberg uncertainty and the weak force would constrain it.

## *5.3 Most of the Usual Expression of Mixing is Preserved*

There really is no need for "mass eigenstates" in the revised neutrino oscillation theory; they are included solely for compatibility with the usual theory, allowing the voluminous published data in the mixing-angle vs. $\Delta m^2$ phase space to remain meaningful.

The **revised neutrino** theory need not say anything about the Higgs boson (if it exists); however, there is no reason a propagating neutrino should not be associated with a global Higgs field, in which it would interact at most virtually, or with a local Higgs particle with which it might interact in some arbitrary way.

The **revised neutrino** described above would:

1. Optionally interact with the mass differences and flavor determined by the usual oscillation theory;
2. Always display the total energy and momentum of its initial state;
3. Conserve mass during propagation (gravitationally and between initial and final interactions);
4. Avoid the question of a reaction "mass field" to quantize eigenvalues of mass, because the new eigenvalues would represent quirks in the weak field of the revised neutrino itself;
5. Allow for matter-dependent interactions (MSW) or other elaborations of vacuum oscillations;
6. Admit of the possibility that masses of different neutrino flavors might be different, definite, or mixed;
7. Admit of the possibility that the neutrino individual masses, or their expectancies, might be found to be equal; and,
8. Disallow the possibility of stale neutrinos.

## *5.4 The Cost of Complexity*

This presentation shows that any of the usual theories for neutrino oscillations has to be more complicated than expected. Therefore, it somewhat weakens the argument that neutrinos might be massive. The argument for neutrino mass so far has been supported solely by the data interpretable as evidence of neutrino flavor oscillations.

The usual oscillation theory in effect projects shadows of the quarks in a hadron onto a screen at infinite distance, thus amplifying the hypothetical effect of the postulated tiny masses of the images--now interpreted as shadows of neutrinos. The amplification seems to be a classical error which disregards fundamental quantum limitations. We consider the question of the mass of the neutrino still to be an open one.

# 6. References

Note: For references marked "*UT-kek*", visit the University of Tokyo at: `http://neutrino.kek.jp` (or, use `http://www-lib.kek.jp/KISS/kiss_prepri.html`). For references marked "*arXiv*", visit Los Alamos National Laboratory at: `http://xxx.lanl.gov`.

## 7. Acknowledgements